\documentclass{ceurart}

\usepackage{booktabs}
\usepackage{array}
\usepackage{amsmath,amssymb}
\usepackage{microtype}

\hypersetup{
  pdftitle={Tie Handling Is Part of the Evaluation Protocol: An Order-Invariance Audit for Tie-Heavy Recommender Scores},
  pdfauthor={Chengkun Guo, Han Chen, Steven Zhu, Yingrui Li},
  pdfkeywords={recommender systems, offline evaluation, score ties, reproducibility}
}

\makeatletter
\patchcmd{\@copyrightLine}{\textcopyright\ \@copyrightyear\ }{}{}{
  \PackageError{frame-camera-ready}{CEURART copyright prefix changed}
    {Check the copyright line before submitting.}
}
\makeatother

\NewCommandCopy{\frameOriginalXmpdatawrite}{\xmpdatawrite}
\RenewDocumentCommand{\xmpdatawrite}{m m}{%
  \ifstrequal{#1}{Copyright}
    {\frameOriginalXmpdatawrite{Copyright}{Copyright (c) 2026 for this paper by its authors. Use permitted under Creative Commons License Attribution 4.0 International (CC BY 4.0).}}
    {\frameOriginalXmpdatawrite{#1}{#2}}%
}

\begin{document}

\copyrightyear{2026}
\copyrightclause{Copyright © 2026 for this paper by its authors. Use permitted under Creative Commons License Attribution 4.0 International (CC BY 4.0).}
\conference{FRAME'26: Methodology First - Rethinking Research Assessment in RecSys Workshop, September 28, 2026, Minneapolis, Minnesota, USA}

\title{Tie Handling Is Part of the Evaluation Protocol: An Order-Invariance Audit for Tie-Heavy Recommender Scores}

\author[1]{Chengkun Guo}[email=guochk@gmail.com]
\author[2]{Han Chen}[email=hc414t@gmail.com]
\cormark[1]
\author[3]{Yilin Zhu}[email=yilin.zhu2@gmail.com]
\author[2]{Yingrui Li}[email=el516y@gmail.com]
\address[1]{Independent Researcher, Bethlehem, Pennsylvania, USA}
\address[2]{Independent Researcher, Washington, DC, USA}
\address[3]{Independent Researcher, Seattle, Washington, USA}
\cortext[1]{Corresponding author.}

\begin{abstract}
Offline top-$k$ evaluation often ranks one held-out relevant item together with sampled negatives. When several candidates receive exactly the same score, the tie-breaking rule becomes part of the ranking. A common implementation stores the relevant item first and then applies a stable sort, which preserves input order among equal scores; the relevant item therefore wins every tie. We call an evaluator \emph{row-order invariant} when permuting the input candidates without changing their identities, labels, or scores leaves the final ranking unchanged. We audit this property by holding candidates and scores fixed and changing only the tie-breaking rule. On 30,000 Amazon Beauty \& Personal Care rows, NDCG@10 for a rating-weighted attribute-overlap score is 0.85 under input-order tie-breaking. A deterministic hash tie-break based on user and item IDs lowers it to 0.17. The exact expectation under uniform random tie-breaking closely matches the mean over 100 independent hash seeds, while a residualized attribute score with few exact ties is nearly unchanged. MovieLens Tag Genome shows the same pattern for an attribute-overlap score, whereas item popularity is nearly unchanged. We derive expected Hit Rate and NDCG at cutoff $k$ when the relevant item is randomly ordered among candidates with the same score, and we provide a practical reporting checklist. The same issue can occur in sampled or full-catalog evaluation whenever exact ties affect top-$k$ membership or rank.
\end{abstract}


\maketitle

\section{Introduction}

Offline recommender results depend on the complete evaluation protocol, not only on the metric formula. Relevant choices include the data split, candidate set, negative-sampling procedure, relevance definition, ranking rule, cutoff, and aggregation method \citep{herlocker2004evaluating,cremonesi2010performance,canamares2020offline,canamares2020target,zangerle2022evaluating}. One choice is often left implicit: how are candidates ordered when their scores are exactly equal?

Consider an evaluation row that stores the held-out relevant item first, followed by sampled negatives. If all candidates receive score zero and the evaluator applies a stable descending sort, the list does not move: the relevant item is ranked first even though the score did not distinguish it from any negative. Reordering the same candidates before sorting can produce a different metric without changing the model output.

We study this dependency through \emph{row-order invariance}. An evaluator is row-order invariant if every permutation of the input candidates that preserves item identities, relevance labels, and scores produces the same final ranking. To test the property, we hold each candidate set and its scores fixed and change only the rule used to order exact ties. Thus, any resulting metric difference is caused by tie handling rather than training, candidate sampling, or scoring.

Prior work has already shown that tie-breaking can affect information-retrieval and recommender-system evaluation \citep{mcsherry2008computing,cabanac2010tiebreaking,bellogin2013empirical,lin2019impact}. Our focus is a specific systematic interaction in recommender evaluation: a relevant item placed at a fixed position, a tie-heavy score, and a stable sort that reuses input order as an undeclared secondary ranking criterion.

This paper makes three contributions. First, it formalizes row-order invariance and derives the expected Hit Rate and NDCG when the relevant item is uniformly placed within an exact-score tie. Second, it compares input-order and deterministic hash tie-breaking in both domains; for Amazon, it also reports the exact expected metric under uniform random tie-breaking and results from 100 independent hash seeds. Third, it gives a six-part reporting checklist covering the candidate set, candidate order before sorting, tie prevalence, tie-breaking, permutation tests, and sensitivity analysis.

\section{Related Evaluation Methodology}

Sampled and full-catalog evaluation define different candidate sets and can produce different metric values or system orderings \citep{krichene2020sampled,canamares2020target}. Our analysis begins after that candidate set has been declared. It asks whether the evaluator gives the same result when the same candidates and scores are presented in a different order.

Tied-score evaluation has a substantial history in information retrieval. McSherry and Najork derive efficient metric computation in the presence of ties \citep{mcsherry2008computing}. Cabanac et al. show that an uncontrolled document-identifier tie-break can change TREC scores \citep{cabanac2010tiebreaking}, and Lin and Yang study the effect of ties on repeatability across ranking implementations \citep{lin2019impact}. Tie-breaking has also been examined in recommender comparisons \citep{bellogin2013empirical}. We build on this literature by isolating the interaction among relevant-item-first rows, stable sorting, and exact ties while holding candidate rows and scores fixed.

Candidate order also matters in listwise large-language-model ranking. Recent work studies order sensitivity and position-invariant listwise reranking \citep{chao2024betterranker,bito2026onepass}. In those settings, changing candidate presentation can change the model's output scores or ranking. Our setting is downstream: scores are already fixed, and only the evaluator's treatment of equal scores changes.

Position bias in logged feedback is a different problem \citep{joachims2017unbiased,chen2023bias,dzhoha2024reducing}. It acts during data collection and learning and can amplify item popularity through feedback loops, whereas the audit here acts after candidate scores have been produced.

\section{When Tie-Breaking Changes the Metric}

\subsection{Candidate rows and exact-score ties}

For a user $u$, let $C_u=(i_1,\ldots,i_m)$ be the ordered candidate list supplied to the evaluator, let $s_u(i)$ be the score of candidate $i$, and let $y_u(i)\in\{0,1\}$ indicate relevance. The experiments use one relevant item and $m=31$ candidates per row, but the definitions are general.

After sorting by descending score, a \emph{tie block} is a maximal consecutive group of candidates with the same score. A tie block \emph{crosses the top-$k$ boundary} when some of its items are above rank $k$ and others are below it. In that case, the tie-breaking rule can determine which items appear in the top-$k$ ranked list. Even when the entire tie block is within the top $k$, the tie-breaking rule can still change position-sensitive metrics such as NDCG.

A \emph{stable sort by score} alone is not row-order invariant because equal-score candidates retain their input positions. A deterministic \emph{hash tie-break} is row-order invariant when it sorts equal-score candidates by a reproducible key computed from a disclosed seed, user ID, and item ID, using item ID as the final fallback if two candidates receive the same key. The key must not depend on the candidate's column or storage position.

\subsection{Expected metrics under uniform random tie-breaking}

Suppose $a$ candidates have scores strictly larger than the relevant item, and the relevant item belongs to a tie block of size $t$. When the relevant item is first in the input row, a \emph{stable sort} gives it rank $a+1$. Under \emph{uniform random ordering} within the tie block, its rank is uniform on $\{a+1,\ldots,a+t\}$. With one relevant item per row,
\begin{align}
\mathbb{E}[\hit@k]
  &= \frac{\max\{0,\min(t,k-a)\}}{t}, \label{eq:expected-hit}\\
\mathbb{E}[\ndcg@k]
  &= \frac{1}{t}\sum_{r=a+1}^{\min(a+t,k)}\frac{1}{\log_2(r+1)}. \label{eq:expected-ndcg}
\end{align}

With one relevant item (the positive) and 30 negatives all tied ($a=0,t=31$), placing the relevant item first gives Hit@10 and NDCG@10 equal to 1. \emph{Uniform random tie-breaking} gives expected Hit@10 $=10/31=0.3226$ and expected NDCG@10 $=0.1466$. The difference follows mechanically from the tie block; it is not evidence that the score identified the relevant item.

Uniformly permuting the complete candidate row independently of relevance before applying a stable sort is equivalent, in expectation, to {uniform random ordering} within each tie block. The metric from a single permutation is therefore an unbiased estimate of the expectation above, but it remains noisy and can change from run to run. Repeated random tie-breaking or direct analytic averaging gives a clearer summary.

\subsection{Practical tie-breaking choices}

We compare four ways to handle exact ties:

\begin{enumerate}
\item \textbf{Input-order tie-breaking.} Sort by score and keep the original candidate order when scores are equal. The result is deterministic but can depend on row construction.
\item \textbf{Deterministic hash tie-breaking.} Within each tie, sort by a reproducible key computed from a disclosed seed, user ID, and item ID, using item ID as the final fallback. This gives one repeatable ranking independent of input order.
\item \textbf{Repeated random tie-breaking.} Break ties with different random seeds and report the metric's mean and variation across runs. This estimates the result under random tie-breaking.
\item \textbf{Exact expectation under uniform random tie-breaking.} Compute the average metric over all possible positions of the relevant item within its tie block. In the one-relevant-item setting, Equations~\ref{eq:expected-hit} and~\ref{eq:expected-ndcg} give this value exactly.
\end{enumerate}

The first method uses input order; the other three do not. Deterministic hashing selects one repeatable ranking. Repeated random tie-breaking estimates the average over random tie orders, while the analytic method computes that average exactly in the one-relevant-item setting.

\section{Audit Design}

\subsection{Candidate sets and scores}

Table~\ref{tab:setup} summarizes the two audits. Every candidate row stores the held-out relevant item first and then 30 sampled negatives before sorting. The candidate-sampling seed is 42 in both domains.

\begin{table}[htbp]
\caption{Audit setup. The Amazon audit uses a deterministic sample of 30,000 rows from 2,073,945 held-out interactions; MovieLens uses 10,000 test users. The last score listed for each dataset is a diagnostic with little sensitivity to tie-breaking; the Amazon and MovieLens diagnostic scores are constructed differently.}
\label{tab:setup}
\centering
\small
\begin{tabular}{>{\raggedright\arraybackslash}p{0.19\linewidth}>{\raggedright\arraybackslash}p{0.28\linewidth}>{\raggedright\arraybackslash}p{0.43\linewidth}}
\toprule
Dataset & Candidate set & Audited scores \\
\midrule
Amazon Beauty \& Personal Care & One relevant item and 30 sampled negatives; 389,478 catalog items & Rating-weighted attribute overlap; centered attribute overlap; residualized attribute score \\
MovieLens 25M + Tag Genome & One relevant item and 30 unique sampled negatives & Unweighted tag-attribute overlap; centered tag overlap; item popularity \\
\bottomrule
\end{tabular}
\end{table}

\paragraph{Amazon Beauty \& Personal Care.}
We use Amazon Reviews 2023 \citep{hou2024bridging}, split chronologically within each user. Negatives exclude items in the user's training and validation history, but the sampler does not enforce uniqueness among the 30 negatives in a row. Three binary lexical proxies extracted from product metadata are used: \emph{fragrance-free}, \emph{cruelty-free}, and \emph{sulfate-free}. Let $H_u$ denote user $u$'s training history, and let $A_i\in\{0,1\}^3$ be item $i$'s attribute vector. For interaction $r\in H_u$, let $i_r$ be its item and let $w_r=\operatorname{clip}((\mathrm{rating}_r-1)/4,0,1)$ be its rating weight. The score named \texttt{raw\_count} in the artifact is
\[
 s_{\mathrm{overlap}}(u,i)=\left(\sum_{r\in H_u} w_r A_{i_r}\right)^{\!\top} A_i.
\]
It measures the rating-weighted overlap between the candidate's attributes and attributes appearing in the user's history. Interactions with higher ratings receive larger weights. Simply counting the number of historical items would be constant across all candidates for a user and therefore could not rank them. The centered variant uses $A_i-\bar A$ for both historical items and candidates, where $\bar A$ is the catalog-wide mean attribute vector.

We include the Amazon residualized score because it has few exact ties and therefore provides a stability check when the primary score already determines the order. The score removes two predictable components before computing the same type of user--candidate dot product. For items, let $X_i$ be the product-metadata text after removing predefined phrases that directly indicate the target attribute. The residual $R_i=A_i-\hat p(X_i)$ subtracts the attribute probability predicted from $X_i$; residuals are set to zero where $\hat p(X_i)>0.95$. For interactions, three-fold cross-fitting predicts each training weight $w_r$ from a smoothed user baseline and user-specific effects for the item's store, category, and price groups, without using the row being predicted. Let $\hat c_r$ denote this out-of-fold prediction for interaction $r$. The user profile is
\[
 \sum_{r\in H_u}\bigl(w_r-\hat c_r\bigr)R_{i_r},
\]
and the candidate score is its dot product with $R_i$.

\paragraph{MovieLens 25M.}
Ratings of at least four stars are treated as positive \citep{harper2015movielens}. Six Tag Genome concepts are thresholded at 0.7 to form binary item attributes: action, classic, visually appealing, based on a book, violence, and sci-fi \citep{vig2012taggenome}. The attribute-overlap score counts these attributes over qualifying training interactions and takes the dot product with the candidate's attribute vector. We also evaluate its centered version and item popularity. We include item popularity because its relevant-item ranks are nearly unchanged across the two tie-breaking rules.

\subsection{Tie-breaking implementations and metrics}

The deterministic rule hashes a fixed seed, user ID, and item ID into an unsigned 64-bit integer, then uses item ID as a final fallback. The Amazon seed is 20260316 and the MovieLens seed is 20260318. The exact mixer and tests are included in the artifact.

The original Amazon evaluator converted the 64-bit hash to 32-bit floating point before sorting. The reference implementation keeps the integer hash. On the reconstructed Amazon rows, the two implementations produce identical aggregate metrics. One row contains a collision between distinct item IDs after float32 conversion; it changes the internal order of two tied candidates but not top-10 membership or the relevant item's rank. The integer implementation remains unchanged when candidate order is permuted.

We report NDCG@10 and Hit Rate@10, the fraction of rows whose relevant item appears in the top 10. With one relevant item per row, Hit@10 equals Recall@10. We also measure the fraction of rows containing any exact tie, the fraction with a tie crossing the top-10 boundary, and the fraction in which the relevant item belongs to a tie.

\subsection{Reproducibility}

For Amazon, preserved historical inputs allow deterministic reconstruction of the 30,000 audited rows and their candidate-level scores. The reconstructed arrays reproduce all 84 aggregate values from the original run exactly, with maximum absolute difference zero. We use those arrays for the integer-hash, analytic-expectation, 100-seed, collision, and permutation analyses.

For MovieLens, we report the aggregate values unchanged from the accepted version. The paired aggregate result file was retained, but the row-level candidate and score arrays were not, so the additional row-level analyses are reported only for Amazon.

The accompanying artifact provides the evaluator, reconstructed Amazon rows and scores, tests, configurations, and provenance records: \href{https://github.com/Guoo123/tie-evaluation-audit}{github.com/Guoo123/tie-evaluation-audit}.

\section{Results}

\subsection{Cross-domain sensitivity to input order}

Table~\ref{tab:crossdomain} reports the main controlled comparison. Candidate identities, labels, sampled negatives, primary scores, and metric code are fixed within each row; only exact-score tie-breaking changes.

\begin{table}[htbp]
\caption{NDCG@10 under input-order and deterministic hash tie-breaking. For Amazon, the integer implementation yields the same aggregate values as the original float32 implementation.}
\label{tab:crossdomain}
\centering
\small
\begin{tabular}{llrrr}
\toprule
Dataset & Score & Input order & Hash tie-break & Change \\
\midrule
Amazon Beauty & Weighted attribute overlap & 0.8474 & 0.1702 & $-0.6772$ \\
Amazon Beauty & Centered attribute overlap & 0.7103 & 0.1627 & $-0.5476$ \\
Amazon Beauty & Residualized attribute score & 0.1689 & 0.1685 & $-0.0004$ \\
\midrule
MovieLens & Tag-attribute overlap & 0.8380 & 0.2332 & $-0.6048$ \\
MovieLens & Centered tag overlap & 0.8406 & 0.2341 & $-0.6065$ \\
MovieLens & Item popularity & 0.5905 & 0.5905 & $-2.2\times10^{-5}$ \\
\bottomrule
\end{tabular}
\end{table}

For the Amazon weighted attribute-overlap score, every audited row contains an exact tie, 99.94\% contain a tie crossing the top-10 boundary, and the relevant item belongs to a tie in 98.84\% of rows. NDCG@10 falls from 0.8474 to 0.1702, while Hit@10 falls from 0.9997 to 0.3526. The centered score is similarly sensitive. The residualized attribute score has an exact tie in only 0.39\% of rows and changes by 0.0004 in NDCG@10.

MovieLens shows the same aggregate pattern in a different item domain. NDCG@10 for tag-attribute overlap changes from 0.8380 to 0.2332, and Hit@10 changes from 0.9390 to 0.4358. The centered score behaves similarly. Item popularity is unchanged to four decimals, showing that changing tie-breaking has little effect when ties do not change the relevant rank.

\subsection{Deterministic, randomized, and expected tie-breaking}

Table~\ref{tab:policies} compares the three input-position-independent treatments alongside the input-order reference on the reconstructed Amazon rows. The exact expected metric under uniform random tie-breaking and the mean over 100 independent hash seeds agree closely. The deterministic hash selects one reproducible ordering and is therefore not required to equal that expectation, although it is close in these data.

\begin{table}[htbp]
\caption{Amazon NDCG@10 under four tie treatments. ``Expected over ties'' analytically averages over all positions of the relevant item within each tie block under uniform random tie-breaking. ``100 hashes'' reports mean $\pm$ standard deviation over independent hash seeds.}
\label{tab:policies}
\centering
\small
\begin{tabular}{lrrrr}
\toprule
Score & Input order & Hash tie-break & Expected over ties & 100 hashes \\
\midrule
Weighted attribute overlap & 0.8474 & 0.1702 & 0.1693 & $0.1694\pm0.0012$ \\
Centered attribute overlap & 0.7103 & 0.1627 & 0.1614 & $0.1615\pm0.0011$ \\
Residualized attribute score & 0.1689 & 0.1685 & 0.1685 & $0.16851\pm0.00003$ \\
\bottomrule
\end{tabular}
\end{table}

The comparison separates two practical goals. Deterministic hash tie-breaking gives a repeatable ranking suitable for a benchmark implementation. The expected metric over ties reports the average over all unresolved positions inside a tie block. Repeated hashes provide a direct numerical check of the analytic result and reveal seed-to-seed variation.

\section{Reporting Tie-Heavy Evaluation}

Tie-breaking can change which items enter the top-\(k\) list and where relevant items rank within it. Ties across the cutoff can affect Hit@\(k\), while ties entirely within the top \(k\) can still affect NDCG@\(k\). The following checklist describes what to report about candidate construction and tie handling.

\begin{enumerate}
\item \textbf{Candidate set.} State whether evaluation is full-catalog or sampled, the number of candidates, the negative-sampling distribution, exclusions, replacement or deduplication behavior, and seeds.
\item \textbf{Candidate order before sorting.} State whether the relevant item occupies a fixed position and whether rows are independently shuffled.
\item \textbf{Tie prevalence.} Report the fraction of rows with any exact tie, with the relevant item in a tie, and with a tie crossing the top-$k$ boundary, separately for each score.
\item \textbf{Tie-breaking rule.} State whether ties use input order, deterministic hash ordering, repeated random ordering, or expected metrics over tie orders. For deterministic or repeated hashes, disclose the inputs, seed policy, and collision fallback.
\item \textbf{Permutation test.} Permute each fixed candidate row and verify that an implementation claimed to be independent of input position returns the same ranking or the same expected metric.
\item \textbf{Sensitivity analysis.} When ties can affect top-\(k\) membership or the positions of relevant items within the top \(k\), report the metric under another tie-breaking rule. Alternatively, report its exact expectation under uniform random tie-breaking. For repeated random tie-breaking, also report variation across runs.
\end{enumerate}

A clean way to remove the systematic relevant-item-first advantage is to permute the entire candidate row independently of relevance before applying a stable sort. This makes the relevant item's position within its tie block uniform in expectation. Each permutation is one random realization, not a fixed invariant ranking. Moving only the relevant item to a randomly selected row position while leaving the negatives in their original order does not generally make its position within the tie block uniform. Full-catalog evaluation can remove the specific relevant-item-first sampled-row construction when catalog order is independent of relevance, but it does not eliminate exact-score ties. A stable sort can still use catalog or storage order as a secondary ranking rule, and a tie crossing the cutoff can still change membership or rank.

\section{Limitations}

The empirical audits cover two public datasets, rows with one relevant item, and 30 sampled negatives. They demonstrate the mechanism and its magnitude for discrete attribute-based scores, but they do not estimate how often the problem occurs across recommender benchmarks or production systems. Continuous rankers may have few exact ties, while quantization, coarse features, fallback scores, or count-based components may create many.

The Amazon attributes are lexical metadata proxies, and the score families are evaluated as components rather than as complete production recommenders. The purpose is to isolate evaluator behavior, not to compare recommendation algorithms. The residualized Amazon score and MovieLens popularity were included to show what happens when tie-breaking rarely changes the relevant item's rank. They serve this comparison purpose but are constructed differently.

Full-catalog evaluation is not examined empirically here. In a fixed full catalog, equal scores still require an explicit ordering or averaging rule. The MovieLens results are limited to the aggregate comparison reported in the accepted version because the corresponding row-level candidate and score arrays were not retained.

\section{Conclusion}

Exact-score ties leave part of a ranking unresolved. If an evaluator resolves them with candidate input order, row construction can enter the metric as an undeclared ranking feature. Holding candidates and scores fixed reveals the effect directly: tie-heavy Amazon and MovieLens scores change by roughly 0.55--0.68 NDCG@10 under deterministic hash tie-breaking, while the residualized Amazon score and MovieLens popularity are nearly unchanged. Deterministic hash ordering provides repeatability; analytic or repeated random tie-breaking describes the average over unresolved orders. Reporting candidate construction, tie prevalence, tie-breaking, and a permutation check makes the evaluated quantity clear and reproducible.

\section*{Declaration on Generative AI}
During the preparation of this work, the authors used OpenAI ChatGPT and Anthropic Claude for grammar and spelling checks, paraphrasing, structural suggestions, LaTeX and CEURART formatting, and implementation assistance in drafting, debugging, and documenting portions of the experimental and analysis code. The research questions, methodology, experimental decisions, result interpretation, and conclusions were determined by the authors. The authors reviewed and edited all affected material, ran and inspected the experiments, verified the text, calculations, citations, tables, and code, and take full responsibility for the publication's content.

\bibliography{references}

\end{document}